\documentclass[12pt]{article}

\usepackage{amsmath}
\usepackage{amssymb}
\usepackage{amsfonts}
\usepackage{graphicx}

\usepackage{latexsym}
\usepackage{color,soul} % Para colocar fundo colorid\o, highlight
\usepackage{cite} % Para agrupar referências consecutivas (ex: [1-3] ao invés de [1,2,3])
\DeclareMathOperator{\cx}{\square}
\def\beq{\begin{eqnarray}}
\def\eeq{\end{eqnarray}}
\def\ln{\,\mbox{ln}}

\newcommand{\nnn}{\nonumber}
\newcommand{\n}[1]{\label{#1}}

\def\al{\alpha}
\def\be{\beta}

\def\ga{\gamma}

\def\ep{\epsilon}

\def\ka{\kappa}

\def\pa{\partial}

\def\si{\sigma}
\def\om{\omega}
\def\ph{\varphi}

\def\th{\theta}

\def\De{\Delta}

\begin{document}

%%%%%%%%%%%%%%%%%%%%%%%%%%%%%
\begin{center}

{\Large
Light bending in \ $F\left[g(\cx)R\right]$ \ 
extended gravity theories
}
\vskip 8mm

{\bf
Breno L. Giacchini$^{a}$,
\ \ and \ \
Ilya L. Shapiro$^{b,c,d}$}

\end{center}
%%%%%%%%%%%%%%%%%%%%%%%%%%%%%
\vskip 1mm

%%%%%%%%%%%%%%%%%%%%%%%%%%%%%
\begin{center}
{\sl
(a) \ Centro Brasileiro de Pesquisas F\'{\i}sicas
\\
Rua Dr. Xavier Sigaud 150, Urca, 22290-180, Rio de Janeiro, RJ, Brazil
\vskip 2mm

(b) \ Departamento de F\'{\i}sica, ICE, Universidade Federal de Juiz de Fora
\\
Campus Universit\'{a}rio - Juiz de Fora, 36036-900, MG, Brazil
\vskip 2mm

(c) \ Tomsk State Pedagogical University, Tomsk, 634041, Russia
\vskip 2mm

(d) \ Tomsk State University, Tomsk, 634050, Russia
}
\vskip 3mm

%%%%%%%%%%%%%%%%%%%%%%%%%%%%
{\sl E-mails:
\ \
breno@cbpf.br,
 \quad
shapiro@fisica.ufjf.br}

\end{center}
%%%%%%%%%%%%%%%%%%%%%%%%%%%%%
\vskip 6mm

\begin{quotation}
\noindent
\textbf{Abstract.} We show that
in the weak field limit the light deflection alone cannot distinguish
between different $\,R + F[g(\cx)R]$ models of gravity, where $F$
and $g$ are arbitrary functions. This does not imply, however, that
in all these theories an observer will see the same deflection angle.
Owed to the need to calibrate the Newton constant, the deflection
angle may be model-dependent after all necessary types of
measurements are taken into account.
\vskip 3mm

%%%%%%%%%%%%%%%%%%%%%%%%%%%%%%%%%
\vskip 2mm

PACS: $\,$
%04.62.+v,	 %%%%%% Quantum fields in curved spacetime
04.20.-q,     %%%%%% Classical general relativity
04.50.Kd 	 %%%%%% Modified theories of gravity
\vskip 2mm

\end{quotation}

%%%%%%%%%%%%%%%%%%%%%%%%%%%%%%%%%%%
%%%%%%%%%%%%%%%%%%%%%%%%%%%%%%%%%%%
\section{Introduction}
\label{S1}

Modified gravity theories attract a lot of attention nowadays, this
concerns in particular the models constructed with the scalar
curvature and its covariant derivatives. In this Letter we discuss the
deflection of light in the linear regime in a generic model of this
type. This issue has already been discussed in the literature using
a variety of approaches applied to specific
models~\cite{Pechlaner-Sexl,Accioly98,
Capozziello06,Clifton,Capozziello11,Berry-Gair,Stabile&Stabile,
Accioly15,3rd_order,ABS}. In the various references one can
meet different conclusions. In some works the statement is that
the bending of light occurs exactly in the same way as in general
relativity (GR), while in other works it is claimed that the light
bending effect can be used to distinguish between some of such
theories and GR. The aim of the present work is to present a
general discussion on the subject and clarify to which extent the
bending of light, taken alone or together with other observables,
can distinguish among the theories which depend on
an arbitrary function $F(Y)$, where $Y=g(\cx)R$, and $g$ is a
function of the d'Alembert operator.

The action of our interest
is given by the expression
\beq
\label{genaction}
\mathcal{S}_{\rm gen}
\,=\,
\int d^4 x \sqrt{-g}
\Big \lbrace
-\dfrac{2}{\kappa^2} R + F\left[g(\cx)R\right]\Big\rbrace\,,
\eeq
where $\,\ka^2 = 32 \pi G$ and $\,G\,$ is Newton's
constant.
The integrand of the gravitational action should be supplemented by
the Lagrange matter density $\,\mathcal{L}_M$.

Our interest will be concentrated on the light bending by a weak
gravitational field, which means that the background is considered to
be  a small deviation from the flat Minkowski space. This condition
immediately provides significant simplification, as the non-linear
in curvature terms are expected to have small effect on the bending
of light. Then the influence of a non-flat background can be safely
regarded as at least a second-order effect. Therefore, without
losing generality, one can restrict consideration by the actions
which are at most quadratic in scalar curvature, but may be
non-polynomial in derivatives. In this way one arrives at the
simplified action of the form
\beq
\label{actionHDG}
\mathcal{S} = \int d^4 x \sqrt{-g}
\left[
-\dfrac{2}{\kappa^2} R
+ R f(\cx) R
+ \mathcal{L}_M \right] \,,
\eeq
where $f(\cx)$ is a function of  d'Alembert operator, which is
not necessarily analytic.

The theory (\ref{actionHDG})
includes many
interesting particular cases. Let us give a short
list of the models of this type which have been discussed in the
literature.
\vskip 1mm

1) \  Taking \ $f(x)=\text{\it const}.$ \
gives the well-known $R+R^2$ gravity, which is
the basis of the
successful  inflationary model of Starobinsky
\cite{star80,star83}. In this case the bending of light is exactly
like in GR, as has been proved in the paper by Accioly \textit{et al.}
\cite{Accioly98}. Our present consideration can be seen as a
generalization of this work to the general model (\ref{genaction}).
Let us note that this statement concerns only the prediction of light
deflection {\it without} taking other observables into account,
as will be discussed in what follows.
\vskip 1mm

2) \  $f(x)$ is a polynomial function of order $N$. The bending of
light for the particular case of a linear function $\,f(\cx)\,$ has
been explored in the recent work \cite{ABS}. The Ricci-squared
term has also been included into consideration in~\cite{ABS}, so the
theory (\ref{actionHDG}) represents only the part of the model which
is related to scalar curvature. Here we generalize the corresponding
results of \cite{ABS} (see also \cite{Seesaw}) to arbitrary
polynomial or non-polynomial functions.
\vskip 1mm

3) \ The logarithmic form factors $\,\ln \left( \cx/\mu^2\right)\,$ are
typical for the quantum corrections coming from the loops of massless
fields, and correspond to the Minimal Subtraction scheme running
of effective constants. At higher loops there may be higher powers of
the same logarithms. In case of massive fields the expressions for the
gravitational form factors are more complicated \cite{apco}, but
qualitatively are like  $\,\ln \left( [\cx+m^2]/\mu^2\right)\,$ for the
quantum field of mass $m$. At low energy this expression becomes
a constant, which corresponds to the gravitational  version of
Appelquist and Carazzone decoupling theorem.
\vskip 1mm

4) \ The functions $\,f_{-1}(\cx)=\mu \cx^{-1}\,$ and
$\,f_{-2}(\cx)=\mu\cx^{-2}\,$ have been introduced in~\cite{apco}
in order to explain how the renormalization group running of the
Einstein-Hilbert and cosmological terms can look like. The main
source of importance of these terms for cosmological applications
is that under the global scaling the $\,Rf_{-1}(\cx)R$-term behaves
exactly like the Einstein-Hilbert term, and the $\,Rf_{-2}(\cx)R$-term
behaves like the cosmological term. At the same time, since both
terms are non-local, they do not reproduce the Einstein-Hilbert and
cosmological terms exactly, and this leads to their fruitful use in
cosmology, as suggested in \cite{DesWood} for
the $R\ph(\cx^{-1}R)$-type actions and in \cite{Maggiore} for the
$\,R\cx^{-2}R$-term (see further references therein). In view of
this phenomenological success, it would be interesting to know
what is the effect of these terms for the light bending.

In what follows we
show that
for the modified
actions~\eqref{actionHDG} in the weak field limit the deflection of light is not directly
affected by the terms quadratic in scalar curvature. This does not
mean, however, that the predictions for the light bending is the same
in all the models~\eqref{actionHDG}, since the experimentally
measured Keplerian mass of astronomical bodies can differ from
theory to theory.

The paper is organized as follows. In Sec.~\ref{s2} the bending
of light is obtained on the basis of tree-level amplitudes, for the
cases of GR and the models (\ref{actionHDG}). The purely classical,
geometric optics approach is used in Sec.~\ref{s3}, where we also
discuss the correspondence with the Brans-Dicke models and the
role of conformal transformations for the light bending and other
observables. Finally, in  Sec.~\ref{s4} we draw our conclusions.

%%%%%%%%%%%%%%%%%%%%%%%%%%%%%%%%%%%
%%%%%%%%%%%%%%%%%%%%%%%%%%%%%%%%%%%
\section{Tree-level approach}
\label{s2}

The interaction between light and a weak gravitational field can be
explored using the scattering approach by means of the tree-level
Feynman diagrams~\cite{BarkerGuptaHaracz,Sabbata}. Loop
corrections may be included as well into this formalism
\cite{Donoghue,Holstein1,Holstein2,Bai17}. In the present work our
interest is concentrated on the classical behaviour of the theories
(\ref{actionHDG}), and this includes the non-local form factors which
can be attributed to the dressed propagator of gravitational modes.
Hence in this framework we only need to deal with the tree-level
diagrams.
For the sake of consistency we shall start by
briefly reviewing the use of the same method within GR, before going
on in applying it to 
extended models~\eqref{actionHDG}.

%%%%%%%%%%%%%%%%%%%%%%%%%%%%%%%%%
\subsection{Light bending in general relativity}
\label{s2.1}

The calculation for GR can be found in some books (see,
e.g.,~\cite{Scadron,Zee}), but our derivation is carried out in a
slightly different way, such that consequent generalization to the
more complicated model \eqref{actionHDG} is straightforward.

The starting point is
the metric
written as a fluctuation around the Minkowski space, i.e.,
$g_{\mu\nu} = \eta_{\mu\nu} + \kappa h_{\mu\nu}$, with
$\eta_{\mu\nu} = \text{diag}(+1,-1,-1,-1)$. Since gravity couples
to matter via the energy-momentum tensor, in the first order in
$\kappa$ the interaction is described by
\beq
\mathcal{L}_{\text{int}}
\,=\, - \, \frac{\kappa}{2}\, h^{\mu\nu} T_{\mu\nu} ,
\n{LT}
\eeq
where $T^{\mu\nu}$ is the matter energy-momentum tensor in
flat space-time, which defines the interaction vertex. For the
electromagnetic field we have
\beq
T_{\text{em}}^{\mu\nu}
\,=\,  - \,\eta_{\alpha\beta} F^{\mu\alpha}F^{\nu\beta}
\,+\, \frac{1}{4} \eta^{\mu\nu}  F_{\alpha\beta}F^{\alpha\beta},
\eeq
with $F_{\mu\nu} = \partial_\mu A_\nu - \partial_\nu A_\mu$.
We will also need the interaction with a minimal
massive scalar,
\beq
\label{T_scal}
T_{\text{scal}}^{\mu\nu}  \,=\,
\partial^\mu \phi \, \partial^\nu \phi
- \frac{1}{2} \eta^{\mu\nu}
\left(   \partial_\alpha \phi \, \partial^\alpha \phi - M^2 \phi^2 \right) \,.
\eeq

In the context of GR, gravitational interaction is related to the exchange of
only one particle, the graviton, whose propagator can be written as
\beq
\label{propGR}
D_{\mu\nu,\alpha\beta}^{\text{GR}} (k)
\,=\, \frac{ P^{(2)}_{\mu\nu,\al\be}}{k^2}
- \frac{P^{(0-s)}_{\mu\nu,\al\be}}{2k^2}
\eeq
in the momentum-space representation. Here $P^{(2)}_{\mu\nu,\al\be}$
and $P^{(0-s)}_{\mu\nu,\alpha\beta}$ are the spin-2 and spin-0
projectors (see, e.g., \cite{book}) defined through
\beq
P^{(2)}_{\mu\nu,\alpha\beta}
& = & \dfrac{1}{2} \left( \theta_{\mu\alpha}\theta_{\nu\beta}
+ \theta_{\mu\beta}\theta_{\nu\alpha} \right)
- \dfrac{1}{3} \theta_{\mu\nu}\theta_{\alpha\beta},
\\
P^{(0-s)}_{\mu\nu,\alpha\beta}
& = &
\dfrac{1}{3} \theta_{\mu\nu}\theta_{\alpha\beta},
\eeq
where the longitudinal and transverse vector-space projectors are
\beq
\omega_{\mu\nu} = \frac{k_\mu k_\nu}{k^2},
\quad \text{and} \quad
\theta_{\mu\nu} = \eta_{\mu\nu} - \omega_{\mu\nu}.
\eeq
Let us note that the other parts of the gravitational propagator are
irrelevant, because it is going to be directly contracted
with the matter sources in the tree-level approximation. For the same
reason we do not need to worry about the gauge-fixing dependence
for the quantum gravitational field. After going on-shell this
dependence will disappear anyway.

At the tree-level we can model the massive source which produces
the gravitational field by a massive scalar field. Therefore we have
to evaluate the exchange of graviton between such a scalar and a
photon.
The leading contribution to the amplitude $\mathcal{M}$ associated
to this process is of the order $\kappa^2$ and is related via the
LSZ reduction formula to the function
\beq
G_4 = -\frac{\ka^2}{4} \iint d^4 z_1 d^4 z_2
\,
\big\langle 0 \big\vert
T \phi(x) \phi(x') A_\mu(y) A_\nu(y')
T_{\text{scal}}^{\al\be}(z_1)
T_{\text{em}}^{\rho\sigma} (z_2) h_{\alpha\beta}(z_1)
h_{\rho\sigma} (z_2) \big\vert 0 \big\rangle  _\text{c} \, .
\n{G4}
\eeq

Using Wick's theorem, the term
$h_{\alpha\beta}(z_1) h_{\rho\sigma} (z_2)$ yields the graviton
propagator~\eqref{propGR}. The energy-momentum
tensor of the electromagnetic field satisfies the conservation law
$\pa_\al T^{\al\be}_{\text{em}} = 0$ and is traceless,
$T^\al_{\al \,\text{em}} = 0$. Consequently, the terms  proportional
to $\om_{\mu\nu}$ in the graviton propagator do not contribute to
the expression (\ref{G4}), and hence to the scattering amplitude.
We can thus replace the projectors $\th_{\mu\nu}$ by the metric
$\eta_{\mu\nu}$ in the graviton propagator.
From the physical perspective the situation is such that photons
interact only with the spin-2 sector of the propagator.
\\
%%%%%%%%%%%%%%%%%%%%%%%%%%%%
 \begin{figure}[!ht]
 \begin{center}
 \includegraphics[height= 3.6 cm,width= 4.8 cm]{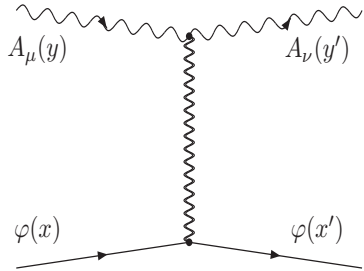}
 \caption{Illustration of graviton exchange between a massive source and
 the photon.}
 \label{F1}
 \end{center}
 \end{figure}
%%%%%%%%%%%%%%%%%%%%%%%%%%%%

The lowest order contribution to the scattering matrix comes from
the tree diagram in~Fig.~\ref{F1},
in which one photon with initial momentum $p$ and
one scalar with momentum $q$ interchange one graviton, resulting
in a final state with momenta $p^\prime$ and $q^\prime$. The
corresponding scattering amplitude is
\beq
\label{AmplitudeGR}
\mathcal{M}
&=&
V_{\mu\nu}^{(\phi)} (q,q^\prime)
\,D^{\mu\nu,\al\be}_{\text{GR}} (k) \,
\,V_{\al\be\,\rho\si}^{(A)}(p,p^\prime)
\,\ep^\rho (\textbf{p}) \,\ep^{*\si} (\textbf{p}^\prime)\, ,
\eeq
where we included the polarization vectors of the photons. The
vertex functions are
\beq
\label{vertexS}
V_{\mu\nu}^{(\phi)} (p,p^\prime) = -\frac{i\ka}{2}
\left[
p_\mu p^\prime_\nu +  p^\prime_\mu p_\nu
+ \eta_{\mu\nu} (M^2 - p \cdot p^\prime ) \right] \,
\eeq
for the scalar, and
\beq
\label{vertexA}
V_{\al\be\,\mu\nu}^{(A)} (p,p^\prime) & = & -\frac{i\kappa}{2}
\big[ p \cdot p^\prime ( \eta_{\mu\alpha}\eta_{\nu\beta}
+ \eta_{\mu\beta}\eta_{\nu\alpha}  - \eta_{\mu\nu}\eta_{\alpha\beta} )
+  \eta_{\alpha\beta} p^\prime_\mu p_\nu
+ \eta_{\mu\nu} ( p_\alpha p^\prime_\beta + p_\beta p^\prime_\alpha )
\nonumber
\\
&& - ( \eta_{\beta\nu} p^\prime_\mu p_\alpha
+ \eta_{\alpha\mu} p^\prime_\beta p_\nu
+ \eta_{\alpha\nu} p^\prime_\mu p_\beta
+ \eta_{\beta\mu} p^\prime_\alpha p_\nu ) \big]  \,
\eeq
for the photon.

As discussed above, in the amplitude~\eqref{AmplitudeGR} one can
replace the propagator $D^{\mu\nu,\alpha\beta}_{\text{GR}}(k)$ by
the expression
\beq
\frac{1}{2k^2}\left( \eta_{\mu\alpha}\eta_{\nu\beta}
+ \eta_{\mu\beta}\eta_{\nu\alpha} \right).
\eeq
Further simplification is achieved by recalling that we are
interested in the deflection of a light ray passing nearby a very
massive object (such as a star or a galaxy), described by the scalar
field. In this regime it is appropriate to work in the large-mass
approximation for the scalar field, that means $\textbf{q}=0$,
$\textbf{q}^\prime=-\textbf{k}$,
$M \gg \vert \textbf{k} \vert$ and $M \gg \vert \textbf{p} \vert$,
and consider that the photon undergoes an elastic scattering
with $\vert \textbf{p} \vert = \vert \textbf{p}^\prime \vert$. For a
vanishing momentum transfer the scalar-scalar-graviton vertex
simplifies to
$V_{\mu\nu}^{(\phi)} = - i \kappa M^2 \delta_\mu^0\delta_{\nu}^0$.
For the remaining vertex, it is useful to work in the Lorentz gauge
for which $\epsilon^0=0$ and
$\boldsymbol{\epsilon} \cdot \textbf{p} = 0$. 
It is then immediate to verify that Eq.~\eqref{AmplitudeGR}
simplifies to
\beq
\label{amplitude}
\mathcal{M} \,=\, - \, \frac{\kappa^2 M^2 E^2}{\vert \textbf{k} \vert^2}
\,\epsilon (\textbf{p}) \cdot \epsilon^{*} (\textbf{p}^\prime) \, .
\eeq
Here $E = p_0 = p_0^\prime$ is the energy of the photon. The
polarization factor can be dropped in the limit of vanishing
momentum transfer (see~\cite{Guadagnini} for further discussion
on helicity effects such as the flip in the case of more general,
rotating sources). Also, we shall insert the normalization factors
$\sqrt{2E_i}$ for each external particle. Recalling that
$\ka^2 = 32 \pi G$, the scattering amplitude then reads
\beq \label{amp.GR}
\mathcal{V} (\textbf{k}) = - \frac{8 \pi G M E}{\vert \textbf{k} \vert^2} \, ,
\eeq
whose Fourier transform gives the interaction potential felt by the photon:
\beq \label{pot.GR}
V(\textbf{r}) = - \frac{2 G M E}{\vert \textbf{r} \vert} \, .
\eeq
As expected this potential is twice the standard Newtonian potential,
which is felt by non-relativistic particles~\cite{BarkerGuptaHaracz}.
This factor two, indeed, took an important role in the development of
GR, since it was missing in the first light bending prediction made by
Einstein based solely on the equivalence principle~\cite{Einstein1911}.
With the potential~\eqref{pot.GR} above and using classical scattering
theory it is possible to compute the deflection angle undergone by a
light ray owed to the weak field produced by a massive
object\footnote{Let us note that one may also evaluate the
unpolarized small-angle scattering cross section from~\eqref{amplitude}
as in Refs.~\cite{Accioly98,Accioly15,ABS,BarkerGuptaHaracz,Sabbata}.
However, as discussed in~\cite{ABS}, this quantum cross section
cannot be generally applied to the deflection by astronomical bodies.}.

%%%%%%%%%%%%%%%%%%%%%%%%%%%%%
\subsection{Light bending in 
extended theories}
\label{s2.2}

The approach described above can be easily adapted to extended
gravity theories, in which the propagator of the gravitational
interaction has a richer structure. For example, in the case of the
action~\eqref{actionHDG} with $f(x)$ being a polynomial function
of degree $N$, there can be $N+1$ scalar particles of mass
$\mu_i \neq 0$ ($i=0,1,\cdots,N$) along with the usual (massless)
graviton. It is possible for the propagator to have
degenerate poles, complex poles, or even to have solely the pole
at $k^2=0$. The crucial point is that all those possibilities stem
from modifications in the scalar sector of the propagator, while the
spin-2 mode propagates in the same way as in GR. Namely, the
propagator of the extended theory~\eqref{actionHDG} can be
written as
\beq
\label{propHDG}
D_{\mu\nu,\alpha\beta}^{\text{ext.}} (k)
&=&
 \frac{P^{(2)}_{\mu\nu,\alpha\beta} }{k^2}
- \frac{P^{(0-s)}_{\mu\nu,\alpha\beta} }{2k^2
\left[1 - 3 \kappa^2 k^2 f(-k^2)\right] } .
\eeq
In order to gain extra degrees of freedom in the tensor sector one
needs to include an $\,R_{\rho\si} {\tilde f}(\cx) R^{\rho\si}$-term
into the action~\cite{highderi}.  In the present work we do not
consider this extension and deal only with the
model~\eqref{actionHDG}.

It is easy to see that, regardless of additional (compared to GR)
degrees of freedom
mediating gravitational interaction in the theory~\eqref{actionHDG},
the lowest order vertex associated to the interaction between
photons and gravitational field is exactly the same as in GR. This
happens because the extra modes are all in the scalar sector, which
interacts through the trace of the energy-momentum tensor, as
discussed in Sec.~\ref{s2.1}. In other words, photons interact only
with the spin-2 (graviton) sector of the propagator, which is the
same as GR, in spite of the $Rf(\cx)R$-terms in the action. This is
the main reason of why the light bending alone cannot distinguish
between original GR and this type of extensions.
One can write for the Feynman amplitude
\beq
\mathcal{M}
& = &
V_{\mu\nu}^{(\phi)} (q,q^\prime) \,D_{\text{ext.}}^{\mu\nu,\al\be} (k)
\, V_{\alpha\beta\rho\sigma}^{(A)}(p,p^\prime) \, \ep^\rho (\textbf{p})
\,\ep^{*\sigma} (\textbf{p}^\prime)\,
\nonumber
\\
& = &
V_{\mu\nu}^{(\phi)}(q,q^\prime)
\left( \frac{\eta^{\mu\alpha}\eta^{\nu\beta}
+ \eta^{\mu\beta}\eta^{\nu\alpha}}{2k^2} \right)
\,V_{\alpha\beta\rho\sigma}^{(A)}(p,p^\prime) \ep^\rho (\textbf{p})
\,\ep^{*\si} (\textbf{p}^\prime)
\nonumber
\\
&=&
V_{\mu\nu}^{(\phi)} (q,q^\prime) \,
D_{\text{GR}}^{\mu\nu,\alpha\beta} (k)
\,V_{\alpha\beta\rho\sigma}^{(A)}(p,p^\prime)
\,\epsilon^\rho (\textbf{p}) \epsilon^{*\sigma} (\textbf{p}^\prime) ,
\eeq
that is the same as in GR~\eqref{amp.GR}. All in all, light bending
alone cannot distinguish among any two theories of the
type~\eqref{actionHDG}. This general result is in agreement with
the calculations
of~\cite{Capozziello11,Berry-Gair,Stabile&Stabile,Accioly15,3rd_order}
for particular models by means of other methods, which will be also
discussed  in Sec.~\ref{s3}.

%%%%%%%%%%%%%%%%%%%%%%%%%%%%%%%%
\subsection{Light massive modes and rescaling of potential}
%% Remark on the interaction with massive particles}
\label{s2.3}

Despite the equality of the interaction potential in the whole class
of theories~\eqref{actionHDG}, the statement that all these theories
predict the same deflection angle for light may be misleading.
The possible effect is due to the difference between the product
of the Newton's constant $G$ and the mass $M$ of the body
which enters into the formulas for the deflection, and the one
which comes from other measurements and/or observations.
Indeed, $G$ is measured in the experiments, e.g., the Cavendish-type
ones. The measurement of the mass $M$ can be achieved
by detailed investigation of the orbits of astronomic bodies. As a
result, in order to predict a deflection angle one has to take into
account not only the interaction potential with relativistic particles,
but also the non-relativistic potential resulting from the gravitational
scattering of massive particles. This quantity can be computed, as
in~\eqref{pot.GR}, as the Fourier transform of the
non-relativistic scattering amplitude
\beq
\mathcal{M}
&=& V_{\mu\nu}^{(\phi_1)}\,
D^{\mu\nu,\alpha\beta}\, V_{\alpha\beta}^{(\phi_2)},
\eeq
where $\phi_1$ and $\phi_2$ are scalar fields with masses $M$ and
$m$. As the energy-momentum tensor for massive scalar fields has
a non-vanishing trace, the non-relativistic potential depends on the
form of the function $f(x)$ and hence on the massive parameters of
the model.

For example, the non-relativistic potential for a polynomial $f(x)$
acquires Yukawa-like correction terms for each massive simple pole
in the
propagator~\cite{Schmidt,Quandt-Schmidt,Giacchini-poles,Newton-high},
\beq
V(\textbf{r})
&=&
- \,\frac{G M m}{\vert \textbf{r} \vert}
\Big( 1 \,+\, \frac{1}{3} \sum_{i=0}^{N}
\prod_{\substack{j = 0 \\ j \neq i}}^N
\frac{\mu_j^2}{\mu_j^2 - \mu_i^2} e^{- \vert \textbf{r} \vert \mu_i}
\Big).
\eeq
If the quantities $\mu_i$
are real and are much
larger than a typical scale $r_\text{lab}^{-1}$ of experimental
measurements of $G$, then the Yukawa terms are suppressed and the
potential reduces to the Newtonian one. On the other hand, if 
the masses of the extra degrees of freedom satisfy
$\mu_i \ll r_\text{lab}^{-1}$ for a given experimental/observational
system, then the exponentials can be approximated by the unit, and
it is possible to show that the potential is roughly $4/3$ of Newton's one~\cite{Quandt-Schmidt,Giacchini-poles}. In this case our laboratory
measurements would actually measure an effective
$G_\text{eff} = \frac{4}{3} G$, thus the predicted bending angle
for a light-ray passing close to the solar limb would be $3/4$ of the
general relativity's prediction.

This simple example shows that the function $f(x)$ can have an
indirect influence on the prediction of deflection angles. The effect
is indirect because it does not concern the bending of light itself,
but the measured value of the product $\,GM$, which appears in
the expression for the bending angle.
In the realistic astrophysical
situations all this concerns only the models with very light,
albeit massive, gravitational degrees of freedom.

%%%%%%%%%%%%%%%%%%%%%%%%%%%%%%%%
%%%%%%%%%%%%%%%%%%%%%%%%%%%%%%%%
%%%%%%%%%%%%%%%%%%%%%%%%%%%%%%%%
\section{Geometric optics and null-geodesic structure}
\label{s3}

Solving the wave equation
may be viewed as a cleaner way of evaluating the bending of
light, based on the assumption that the source of the gravitational
field is classical, not quantum.
As discussed
in~\cite{ABS} 
this procedure avoids possible subtleties
which appear when using the calculations via Feynman diagrams.
For this reason, in this section we describe how the higher-derivative
and non-local extensions~\eqref{actionHDG} manifest in the bending
of light using classical, geometric arguments.

The analogy between the propagation of massless particles in curved
space-time and geometrical optics in a medium can be regarded as one
of the most straightforward methods for describing the gravitational
deflection of light. One can consult, e.g., Refs.~\cite{LightmanLee,FischbachFreeman,PadLivro,Anderson}
for the detailed description of this method. Let us give a brief
survey of the main results.  For a static and spherically symmetric
source the metric is assumed  to have an isotropic form,
\beq
\label{isot_metric}
ds^2
&=&
(1 + 2 \Phi) dt^2
- (1 - 2\Psi) \big[ dr^2 + r^2 \left( d\theta^2
+ \sin^2\theta \, d \phi^2\right) \big],
\eeq
where $\Phi=\Phi(r)$ and $\Psi=\Psi(r)$ only. It is possible to
demonstrate that in this case light propagation in the geometric
optics limit is equivalent to the propagation in the flat space in
a medium with the effective local refractive index
\beq
\label{index}
n(r)
&=&
\sqrt{\frac{1 - 2\Psi(r)}{1 + 2\Phi(r)}}.
\eeq
The deflection angle can be evaluated with Snell-Descartes law
(see~\cite{FischbachFreeman} for an explicit calculation). This
method is also equivalent to the use of Maxwell equations in a
space-time which represents a small fluctuation around Minkowski one~\cite{LightmanLee,PadLivro}.

In the case of GR, to the first order in $G$, the
metric~\eqref{isot_metric} associated to a point source of mass
$M$ is given by the potentials
\beq
\Phi (r) &=& \Psi (r)  \,=\, - \, \frac{G M}{r}  .
\eeq
Due to the smallness of these
quantities, Eq.~(\ref{index}) yields the
effective refractive index in GR,
\beq
\label{n_GR}
n_{\text{GR}}(r)
&=&
1 \,+\, \frac{2GM}{r} \,+\, O(G^2) .
\eeq

Let us derive the effective refractive index for
extended models of gravity \eqref{actionHDG} with
an arbitrary function $f(x)$.
To this end we rewrite the potentials $\Phi$ and $\Psi$ in
terms of the new functions $\Phi_*$ and $\Psi_*$,
\beq
\label{decomposition}
\Phi (r)
&=& V(r) + \Phi_* (r)
, \quad
\Psi (r) = V(r) + \Psi_* (r)
, \quad
\text{with}
\quad
\Delta V = \frac{\ka^2 \rho}{8} .
\eeq
In the last expression $\rho$ is the static density of matter sourcing
the field. It is clear that for GR the equations of motion must yield
$\Phi_* = \Psi_*\equiv 0$. The solution in the
form~\eqref{isot_metric} for the field generated by a static mass
distribution can be obtained from the trace and the (0,0)-component
of the equations of motion, which read
\beq
&&
4\big[1-3h(-\Delta)\big] \Delta (\Phi - 2 \Psi)
\,=\,  \kappa^2 \rho ,
\\
&&
4\big[ h(-\Delta)-1\big] \Delta \Phi - 8 h(-\Delta) \Delta \Psi
=  - \kappa^2 \rho ,
\\
\mbox{where}
&&
h(\cx) = \left[ 1 + 2 \kappa^2 f(\cx) \cx \right]\,.
\eeq
Let us remember that in the static case $\cx$ boils down to $\,-\De$.
Using the \textit{Ansatz}~\eqref{decomposition} it is
straightforward to show that the functions $\Phi_*$ and $\Psi_*$
satisfy the equations
\beq
\frac{1-3h(-\Delta)}{1 - h(-\Delta)}\, \Delta \Phi_*
= \frac{\ka^2 \rho}{8},
\qquad
\frac{1-3h(-\De)}{1 - h(-\De)} \,\De \Psi_*
= - \frac{\ka^2  \rho}{8},
\eeq
from which we assume $\Phi_*(\textbf{r}) = - \Psi_*(\textbf{r})$.

Taking the point-like mass source
$\rho(\textbf{r}) = M \delta^{(3)}(\textbf{r})$, the effective
refractive index is
\beq
n_{\text{ext}}(r)
& = &
\sqrt{\frac{1 - 2V - 2\Psi_*}{1 + 2V + 2\Phi_*}}
\nonumber
\\
& = &
1 - \left( V - \Phi_* \right)
 - \left( V + \Phi_* \right) + O(G^2)
\label{Ref2}
\\
& = & 1 + \dfrac{2 GM}{r} + O(G^2) .
\label{Ref3}
\eeq
The above expression is
the same one which stems from GR~\eqref{n_GR}.  Once again, we
see that the light deflection alone cannot be used to distinguish between
the two theories. This matches and generalizes the previous results of~\cite{Accioly98,Capozziello11,Berry-Gair,Stabile&Stabile,
Accioly15,3rd_order} in the context of the $R+R^2$ gravity and
of~\cite{ABS} for the sixth-order gravity. A quick inspection
of~\eqref{Ref2} reveals that this happens because of the cancellation
of the $R$-extended contribution, which appears with exactly the same
value, but with an opposite sign, in the $h_{00}$ and $h_{11}$
components. However, as it was discussed in the previous section,
in order to define the light bending one has to establish the quantity
$GM$ from a qualitatively different measurement.

To better explain this conclusion one may write the
metric~\eqref{isot_metric} in the PPN-inspired
form~\cite{Will-book}, with
\beq
\label{PPN_metric}
\Phi = - \frac{G_{\text{eff}}M}{r} , \qquad
\Psi = - \frac{\ga G_{\text{eff}} M}{r},
\eeq
being $G_{\text{eff}}$ the quantity measured in a given experiment.
Let us stress that it is not generally possible to cast theories of the
form~\eqref{actionHDG} into the PPN formulation~\cite{Will-book}.
This happens because the non-relativistic limit in these theories may be
non-Newtonian, since the potential is not proportional to $r^{-1}$ only
but may have an additional non-trivial dependence on $r$, as
discussed in \cite{Clifton,Capozziello10,Peri,Will-2012}.
Indeed, the quantities $G_\text{eff}$ and $\gamma$ are actually
functions of $r$. In the phenomenologically relevant models of this
sort one can thus expect a scale for which it is possible to observe
deviations from the Newtonian mechanics, such as anomalous
precession of elliptical orbits~\cite{S2star1,S2star2}. The force law
can be investigated in such a regime and the free parameters of the
model determined, including the gravitational constant $G$.
Nevertheless, it may happen that these effects are not observable at
a certain scale $\bar{r}$ for which $G_\text{eff}(\bar{r})$ and
$\ga(\bar{r})$ are roughly constant and the laws of mechanics are
 close to the ones of Newton. In this scenario one can work with the
 metric in the PPN form above
 (see, e.g.,~\cite{Clifton,Peri,Will-2012}). Then, from the relations
\beq
&&
\Phi = V + \Phi_* = V \Big( 1 + \frac{\Phi_*}{V}\Big)
\nonumber
\\
\mbox{and}
&&
\Psi = V - \Phi_* = \ga V \Big(1+\frac{\Phi_*}{V}\Big)
\nonumber
\eeq
in the vicinity of $\bar{r}$ one reads off
\beq
\label{EffParam}
G_{\text{eff}}=\left( 1 + \frac{\Phi_*}{V}\right) G
\qquad \text{and} \qquad
\gamma = \frac{V-\Phi_*}{V+\Phi_*}.
\eeq

Assuming that \ $\Phi_*(\bar{r}) \ll V(\bar{r})$ \ in a certain
region, it follows that
\beq
\ga(\bar{r}) \,\approx \,  1 - \frac{2 \Phi_*(\bar{r})}{V (\bar{r})}.
\nnn
\eeq
On the other hand, if the condition \ $\Phi_*(\bar{r}) \ll V(\bar{r})$ \
does not hold, the parameter $\ga$ may be very different from the
value $\gamma_{\text{GR}} = 1$ of GR. Then the bending angle
can vary from theory to theory. For instance\footnote{The
derivation of this result can be found in textbooks on GR, for
the terms of Snell-Descartes law see~\cite{FischbachFreeman}.}, from~\eqref{Ref3} the deflection $\theta$ undergone
by a light ray with impact parameter $b$ is
$\theta = 4GMb^{-1}$. In terms of $G_{\text{eff}}$ this result
has the form
\beq
\label{Theta}
\th \,=\, \frac{4G_{\text{eff}}M}{\left( 1 + \Phi_*/V \right) b}
\,=\,
2\left(1+\ga\right) \,\frac{G_{\text{eff}}M}{b} ,
\eeq
with $\gamma$ defined in~\eqref{EffParam}.

Let us consider a simple example. In the case of a constant function
$f(\cx) \equiv (6\alpha)^{-2}$ (that is, $R+R^2$ gravity) with a
point-like mass as a source one has ($m \equiv \alpha\kappa^{-1}$)
\beq
\label{PPN_param}
\Phi_*(r) = - \frac{GM e^{-mr}}{3r},
\qquad
\gamma = \frac{3 - e^{-mr}}{3 + e^{-mr}}.
\eeq
If $m$ is sufficiently large, the Yukawa term has a range
shorter than the scale of the recent torsion-balance experiments.
Then for all practical purposes the quantity $G_\text{eff}$
measured in laboratory is equal to $G$ and $\gamma = 1$. On
the other hand, if the scalar is very light, with a range much larger
than the Solar System scales, then the measured solar mass in
terms of Newtonian mechanics is such that
\beq
(GM)_\text{eff} \,=\, \Big(1 + \frac13\, e^{-m\bar{r}}\Big)
GM \,\approx \,\frac{4GM}{3},
\nnn
\eeq
and $\ga \approx 1/2$. This example shows that one cannot assume
$\Phi_* \ll V$ in general. Inserting $\ga=1/2$ into~\eqref{Theta} one
finds the bending angle $\th = \frac{3}{4}\theta_{\text{GR}}$, in
agreement with~\cite{Pechlaner-Sexl} (see also the example briefly
discussed in Sec.~\ref{s2.3}). The PPN parameter $\gamma$
in~\eqref{PPN_param} was derived in~\cite{Berry-Gair}, where it
was also argued that its correct value should be, instead, $\ga=1$ in
view of the fact that $\Phi + \Psi = 2 V$ as in~\eqref{Ref3}. As we
have shown above there is no conflict between Eqs.~\eqref{Ref3}
and~\eqref{PPN_param} after one takes into account the observable
quantities in~\eqref{Ref3}. The result $\gamma = 1$ was found
in~\cite{Clifton} under the assumption that the free parameter
$m$ is such that the theory does not violate known experimental
results. This is in agreement with our discussion, i.e., considering
that $m$ is such that it may affect the Newtonian force law only
at sub-micrometre scale. Finally, the deflection of light in this model
was investigated also in
Refs.~\cite{Accioly98,Capozziello11,Stabile&Stabile,3rd_order}
with the conclusion that it cannot be used to distinguish between GR
and $R+R^2$ at the leading order in $G$ (see~\cite{3rd_order} for
calculations at the next order). According to the discussion in the
present work this assertion is partially correct, since any prediction
is based on the quantity $GM$ and the measured $(GM)_\text{eff}$
can differ from GR and quadratic gravity.

%%%%%%%%%%%%%%%%%%%%%%%%%
\subsection{Relation to conformal transformation of the background}
\label{s3.1}

The description of the bending of light in terms of the effective
refractive index is closely related to the fact that light rays follow
null-geodesics associated to the metric $g_{\mu\nu}$. In fact,
the statement that the refractive index for the extended
model~\eqref{actionHDG} equals general relativity's one to the
first order in $G$ can be translated to the geodesic structure of
the manifolds in both theories in terms of conformal transformations.
This association has been carried out, for example,
in~\cite{Whitt84,Teyssandier89} for the model $R+R^2$, and
in~\cite{Schmidt,Quandt-Schmidt,Wands} for the more general
model which is a polynomial function of d'Alembertian. Using the
metric~\eqref{isot_metric} parametrised as~\eqref{decomposition}
and with the result $\Psi_* = - \Phi_*$ obtained above it is cursory
to verify that
\beq
\label{conformal}
g_{\mu\nu}^{\text{ext}} \,=\,
( 1 + 2 \Phi_* ) g_{\mu\nu}^{\text{GR}} ,
\eeq%
were only the terms to order $G$ were kept. Here
\beq
g_{\mu\nu}^{\text{GR}}(r) \,=\, \eta_{\mu\nu} + 2 V\delta_{\mu\nu}
\nnn
\eeq
is GR's weak-field metric associated to the same matter configuration
in the de Donder gauge.

Since null-geodesics are conformal invariant, up to $O(G)$
the null-geodesics of GR are also null-geodesics in the extended
model~\eqref{actionHDG}. As a result, light rays undergo the same
gravitational deflection in both theories. On the other hand, time- and
space-like geodesics are not conformally mapped to geodesics, and
massive particles ought to deflect in a different manner.
As mentioned before, the reasoning presented in this section
parallels the description in terms of Maxwell equations on a
curved background.  The conformal transformation
$g_{\mu\nu}^{\text{GR}} \mapsto g_{\mu\nu}^{\text{ext}}$
does not affect the equations of motion of the electromagnetic
field, as we discussed above (see also \cite{Buchdahl}).

The result~\eqref{conformal} turns out to be the linearised version 
of the theorem proved in~\cite{Schmidt} on the relation of the
model~\eqref{actionHDG}, scalar-tensor theories and GR with extra 
scalar fields\footnote{Indeed, it can be shown that the 
higher-derivative theories whose Lagrangian is formed by functions
of $\cx^n R$ (for all $n \in \mathbb{N}$) are dynamically equivalent
to a (multi)scalar-tensor theory, which is then conformal equivalent
to GR with scalar fields~\cite{Wands}.}. By evoking the same 
arguments in terms of conformal transformations (see, e.g., 
\cite{BD1}), it follows that in the generalized Brans-Dicke (BD) 
theory~\cite{BD} in the form
\beq
\label{actionBD}
\mathcal{S}_{\text{BD}} = \int d^4 x \sqrt{-g}
\left[
\phi R + \omega g^{\mu\nu} \partial_\mu \phi \partial_\nu \phi
- U(\phi) + \mathcal{L}_M \right]
\eeq
the non-trivial contribution to the bending of light also comes from
the observable value $(GM)_{\text{eff}}\,$ rather than to a new force
acting on light rays.

The PPN parameter $\gamma$ for massive Brans-Dicke (mBD) theories,
i.e., with $\omega \neq 0$ and $U^{\prime\prime}(\phi) \neq 0$, was
computed in~\cite{Peri,Will-2012,OlmoPRL,OlmoPRD}, with the result
\beq
\ga_{\text{mBD}} = \frac{2\om + 3 - e^{-mr}}{2\om + 3 + e^{-mr}} ,
\n{BD}
\eeq
where $m$ depends on $\omega$ and on the form of $U$. For a suitable
potential $U$ and in the limit $\omega \rightarrow 0$, it is possible
to show that $\gamma_{\text{mBD}}$ above reduces to the expression
in~\eqref{PPN_param}, evaluated for the $R+R^2$ gravity. This is
expected, since this theory is dynamically equivalent to a BD theory
with $\omega=0$ but with a non-trivial potential $U$~\cite{Tey-Tou}.
Light deflection (or the PPN parameter $\gamma$) has been computed
in the scalar-tensor formulation in~\cite{OHanlon,OlmoPRL,OlmoPRD}
with the results consistent to those presented in the previous
sections, as expected. We note that if $m \ll r^{-1}$ (i.e., the scalar
field has a range much larger than the experimental scale) then
$\gamma_{\text{mBD}} \approx (2\omega+2)/(2\omega+4)$, which gives
$\gamma_{\text{mBD}} \approx 1/2$ for the quadratic equivalent model
with $\omega = 0$, in agreement to the example discussed in the last
section.  It is important to notice, however, that this only occurs
for a very light  massive field; in the case $m \gg r^{-1}$ and $\om = 0$
one has $\gamma_{\text{mBD}} \approx 1$. This situation is different
from the massless BD theory, for which $\omega = 0$ implies in
$\ga_{\text{BD}} = 1/2$ independently of the scale.

%%%%%%%%%%%%%%%%%%%%%%%%%%%%%%%%%%
%%%%%%%%%%%%%%%%%%%%%%%%%%%%%%%%%%
%%%%%%%%%%%%%%%%%%%%%%%%%%%%%%%%%%
\section{Conclusions}
\label{s4}

The gravitational deflection of light is one of the main predictions
of GR, which has a growing precision of measurement. It is quite
reasonable to use this effect to test the models of modified gravity,
including higher-derivative and non-local ones. In the literature one
may find different conclusions on the bending of light within the
simplest extension of GR, the $R+R^2$ gravity. A recurrent
statement is that light bending cannot distinguish between this
theory and GR in the linear
regime~\cite{Accioly98,Capozziello11,Berry-Gair,Stabile&Stabile,
Accioly15,3rd_order}. This result directly follows from the fact
that in quadratic gravity light is not subjected to an extra potential
other than the Newtonian one. However, the situation may change
if taking into account what are the true observable quantities in the
theory. In this context  the predictions for the light bending can
differ, as the Keplerian mass of the object responsible for the
deflection is theory-dependent~\cite{Pechlaner-Sexl,
Clifton,Peri,Will-2012,OlmoPRL,OlmoPRD}. Similar results have
been obtained for the bending of light in the sixth-order
gravity~\cite{ABS,Seesaw}.

In the present work we extended such considerations to the whole
family of gravity theories based on the Einstein-Hilbert action
with an extra term $\,R f(\cx) R$. Namely, we presented different
arguments to show that in the weak field limit the deflection of photons is not
\emph{directly} affected by the terms quadratic
in scalar curvature, regardless of the form of the function $f(x)$.
It turns out that this fact is closely related to the conformal invariance
of Maxwell equations. It is worth noticing that in the weak-field
approximation the result applies also to the general models of
the form~\eqref{genaction}.

The form of $f(x)$ can produce an indirect effect on the light
bending through the redefinition of the (measured) Keplerian mass
$M_{\text{eff}}$. In particular, the effective PPN parameter $\ga$
was calculated as a quantity dependent on the function $f(x)$,
assuming that its spatial dependence can be neglected in some
regimes. This derivation confirms that the predictions
of light bending can differ  from theory to theory among the
class~\eqref{actionHDG}, even though they have the same
interaction potential with light (written in terms of $G$).

Finally, it is worth to point out that the introduction of the terms
of the second order in Ricci and/or Riemann tensors---such as
$R_{\mu\nu} g(\cx) R^{\mu\nu}$, for a local or non-local function
$g(x)$---changes the propagator of the spin-2 sector, which yields a
\emph{direct} effect in the light bending~\cite{Accioly15,ABS,Seesaw}.
More general results in this direction will be presented elsewhere.

%%%%%%%%%%%%%%%%%%%%%%%%%%%%%%%%%%
%%%%%%%%%%%%%%%%%%%%%%%%%%%%%%%%%%
\section*{Acknowledgements}
B.L.G. is thankful to CNPq for supporting his Ph.D. project.
I.Sh. is grateful to CNPq, FAPEMIG and ICTP for partial support
of his work.

%%%%%%%%%%%%%%%%%%%%%%%%%%%%%%%%%%

\end{document}